\documentclass[11pt]{article}

\def\includefigs{y}
\normalsize 
\def\cut#1{}

\usepackage{times}
\usepackage{cite}
\usepackage{graphicx}

\title{Colloidal particle motion as a diagnostic of DNA conformational transitions}
\author{Philip C. Nelson\\ Department of Physics and Astronomy,\\ University of  Pennsylvania,\\ Philadelphia, Pennsylvania 19104 USA\\
{\tt nelson@physics.upenn.edu}\\
voice: +1-215-898-7001; fax: +1-215-898-2010}

\def\cocisstar{$\star$}\def\cocisstarstar{$\star\star$}

\newcommand{\vr}{\mathbf{r}}

\newcommand{\tlf}{\tau_{\scriptscriptstyle\rm LF}}
\newcommand{\tlb}{\tau_{\scriptscriptstyle\rm LB}}

\def\rhomax{\rho_{\rm max}}
\def\kbt{k_{\rm B}T}
\def\nMunit{\ensuremath{\mbox{n\textsc m}}}

\def\bpunit{\ensuremath{\mbox{bp}}}
\def\nmunit{\ensuremath{\mbox{nm}}}

\def\sunit{\ensuremath{\mbox{s}}}

\def\exv#1{\langle #1\rangle}

\newcommand{\inv}{^{\raise.15ex\hbox{${\scriptscriptstyle
-}$}\kern-.05em 1}}

\def\pnlabel#1{
\label{#1}}

\def\sref#1{Sect.~\ref{s:#1}}
\def\fref#1{Fig.~\ref{f:#1}}

\def\rref#1{Ref.~\citen{#1}}
\def\rrefs#1{Refs.~\citen{#1}}

\def\yesflag{y}
\def\ifig#1#2#3{\begin{figure}[tb!]
\begin{minipage}[b]{.95\textwidth}
\ifx\includefigs\yesflag
\begin{center}
\includegraphics
{#3}\end{center}\fi
 \end{minipage}
\hfil\par\smallskip \caption{\footnotesize #2 \pnlabel{f:#1}}
\end{figure}%
}
\def\ifigfour#1#2#3{\begin{figure}[tb!]
\begin{minipage}[b]{.95\textwidth}
\ifx\includefigs\yesflag
\begin{center}
\includegraphics[width=4truein]
{#3}\end{center}\fi
 \end{minipage}
\hfil\par\smallskip \caption{\footnotesize #2 \pnlabel{f:#1}}
\end{figure}%
}
\def\ifigab#1#2#3#4{
\begin{figure}[tb!]
\ifx\includefigs\yesflag%
\makebox[1.1\textwidth]{\begin{minipage}[b]{.45\textwidth}
\begin{center}\includegraphics{#3}\end{center}
\end{minipage}\hfill
\begin{minipage}[b]{.45\textwidth}
\begin{center}\includegraphics{#4}\end{center}
\end{minipage}\hfill}
\fi
\par\smallskip
\caption{\footnotesize #2 \pnlabel{f:#1}}
\end{figure}}

\begin{document}

\maketitle
\begin{abstract}
Tethered particle motion is an experimental technique to monitor conformational changes in single molecules of DNA in real time, by observing the position fluctuations of a micrometer-size particle
attached to the DNA. This article reviews some recent work on theoretical problems inherent in the interpretation of TPM experiments, both in equilibrium and dynamical aspects.

\end{abstract}


\section{Introduction}
Molecular biophysics attempts to explain life processes by understanding the behavior of individual macromolecules. Broadly speaking, past work has relied on (a)~traditional light microscopy, which
can discern the actions of individual objects, in physiological conditions and in real time, but is limited to scales bigger than half a micrometer, (b)~electron microscopy and x-ray diffraction,
which resolve fractions of a nanometer but give little information about dynamics, and (c)~electrophysiology, which gives detailed dynamic information but with an indirect readout, applicable mainly
to one class of molecular device. The recent rise of single molecule biophysics has complemented these approaches, in some ways combining their strengths. Working at the single molecule level
allows us to see differences in behavior between different molecules, as well as making it possible to extract detailed kinetic information. 

Tethered particle motion (TPM) is an experimental technique to monitor conformational changes in single molecules of DNA in real time.
In this technique, a large colloidal particle (typically half a
micrometer in diameter) is attached to the end of a DNA molecule of interest (the ``tether''); the other end is fixed to the microscope coverslip, and the resulting motion is passively observed. This
review will outline the basics of the method,  then turn to more recent applications to the kinetics of DNA looping. Noteworthy progress in the past twelve months includes 
\textit{(1)}~First principles prediction of the calibration curve for bead excursion versus tether length;
\textit{(2)}~Quantitative formulas for a new entropic force stretching DNA in the bead-tether configuration;
\textit{(3)}~Direct and indirect real-time observation of the formation of very short repressor- and restriction-enzyme mediated DNA loops; 
\textit{(4)}~Application of hidden Markov modeling to extract loop-formation kinetics from noisy TPM data.

\section{TPM technique\label{s:tt}}
The TPM technique uses the motion of a large, optically resolvable object (a bead) to report on changes in a nanoscale object of interest (a DNA molecule). The change of interest could include
the progress of a processive motor that walks along DNA, as well as loop or kink formation on the DNA as proteins bind to it (\fref{cartoon}). Experimental aspects of implementing the technique, 
including the attachment of the DNA of interest to the mobile bead at one end and the immobile surface at the other, are discussed in the original articles, including
[\citen{scha91a}\cocisstarstar,\citen{yin94a},\citen{finz95a}\cocisstarstar,\citen{poug04a},\citen{blum05a},\citen{nels06a},\citen{broe06a}\cocisstarstar,\citen{vanz06a},\citen{lamb06a
}]. A related scheme to TPM allows an intramolecular reaction (such as DNA looping) to take place with the bead free, then 
applies force to read out the result of the reaction (e.g.\ loop type\ \cite{gemm06a}); this review will focus only on zero-force TPM experiments.
\ifig{cartoon}{A DNA molecule flexibly links a bead to a surface. The motion of the bead's center is observed and tracked, for example as described in \rrefs{zurl06a,nels06a}. In each video frame, the
position vector, usually projected to the $xy$ plane, is found. After drift subtraction, the mean of this position vector defines the anchoring point; the vector $\vr\cut{_{\perp}}$ discussed in this
article is always understood to be the drift-subtracted, projected position, measured relative to this anchoring point. The length of this vector is called $\rho$. In this cartoon, the conformational 
change of interest is loop formation: Regulatory proteins, for example dimers of lambda repressor cI, bind to specific ``operator'' sites on the DNA. A loop forms when two (or more) such dimers 
subsequently bind to each other. A closely related system involves the lactose repressor protein, LacR (also called LacI):  LacR exists in solution as a tetramer, which has two binding sites for DNA. 
Looping occurs when such a tetramer binds first to one, then to the other of two distant operator sites on the DNA. In each case, recent experiments have sought
to determine the values of the time constants $\tlf$ and $\tlb$ for loop formation and breakdown. 
\cut{The figure is simplified; in the actual experiment leading to the data we analyze, the DNA had two sets of three binding sites
(``operators'') for cI repressor protein [\citen{beau07a}\cocissstar], and each operator can at a given moment be occupied or unoccupied. (Cartoon redrawn after 
[\citen{beau07a}\cocisstar{} Fig.~1].)}}{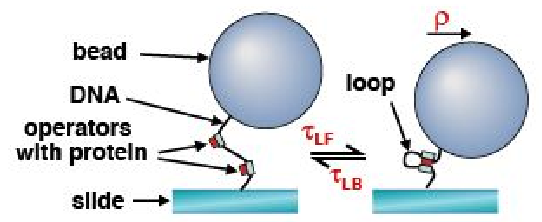}

Once the DNA/bead constructs are created, typically they are attached to the surface with a density small enough to prevent mutual interference, yet large enough that ten or more beads are
simultaneously visible in a microscope frame. The tethered beads are then observed, for example using differential interference contrast microscopy. The observations are typically at video rate, 
either gathering light throughout the entire frame time (typically 33 msec), or (with an electronic shutter or stroboscopic illumination) for some fraction of the frame (for example, 1 msec \cite{nels06a}).
Automated data acquisition software can then track one or more beads in real time, characterizing each bead's motion; again details are available in the 
articles cited above. Some implementations do not track individual trajectories, instead observing the blurred average image of each bead [\citen{finz95a
}\cocisstarstar] (O. K. Wong, M. Guthold, D. A. Erie, J. Gelles, personal communication); this review will focus on 
particle-tracking implementations, e.g.\  \cite{poug04a,nels06a}.

Real-time tracking allows the experimenter to accept or reject individual beads prior to the start of datataking. For example, beads stuck to the surface will display little or no Brownian motion and
can be rejected. During datataking each bead's image is fitted to a standard form to infer its position projected to the microscope focal plane; some experiments
also report on the third component of bead position \cite{blum05a}. Each bead wanders about an average position, its anchoring point; \fref{FIGTIMERHO} shows a typical time series for the bead's
distance from its anchoring point.

\ifig{FIGTIMERHO}{Typical time series of bead positions. DNA constructs of total length $3477\,$ basepairs (\bpunit) were attached at one end to a glass coverslip and at the other to a $480\,\nmunit$ diameter
bead. The vertical axis gives the distance of the bead center from its attachment point, as reported by particle-tracking software, after drift subtraction. The trace shown passed the tests discussed in \rref{nels06a}, for example the
ones that eliminate doubly-tethered beads. The DNA construct contained two sets of three operator sites. The two sets of operators were separated by $2317\,\bpunit$. The system contained cI repressor
protein dimers at concentration 200$\,\nMunit$; repressor proteins bind to the operator sites on the DNA, and to each other, looping the DNA as in \fref{cartoon}.  A sharp transition can be seen from a
regime of no loop formation to one of dynamical loop formation at $\sim$650 \sunit. The
dashed lines represent the two values of $\rhomax$ corresponding respectively to the looped and unlooped states; control data in these two states was observed never to exceed these values. A brief
sticking event, indicated by the inverted triangle, was excised from the data prior to analysis. (Experimental data redrawn after [\citen{beau07a}\cocisstar{} Fig.~2].)
 }{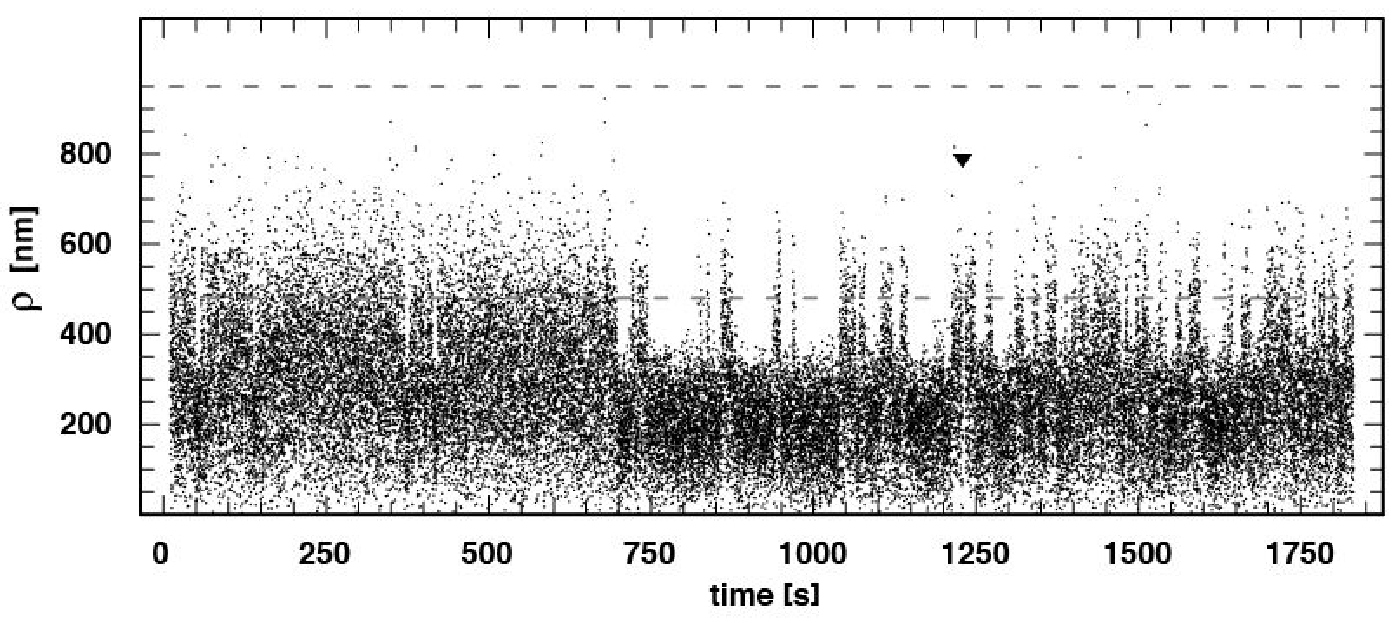}
Typically at least several acceptable beads will survive these initial cuts in a single microscope frame; the experimenter can select those for tracking, then observe their motion 
for an hour or more (\fref{FIGTIMERHO}). This simultaneous acquisition confers some advantages over other single molecule methods: First, it allows a degree of parallel processing, increasing experimental throughput. 
Second, as described later the availability of multiple time series, in time registry, assists in the removal of instrumental vibration and drift. The very long observation times available open the 
possibility of observing very slow kinetics. In contrast, other single molecule measurements such as F\"orster resonance energy transfer (FRET) are limited in duration to the lifetime of a single flurophore.

After datataking, data must generally be be examined for failures of the particle-tracking software: Occasional individual outlier points can be deleted from the time series or replaced by the mean of
flanking points, whereas time series where the tracking software loses a bead altogether must be discarded or truncated. Next, some correction must be applied for the effects of instrumental drift.
Slow drift can be partially removed by low-pass filtering the bead position time series. Most experiments take the additional step of running a sliding window over the data, typically of 4$\,\sunit$
duration, and within each window computing the variance of bead position, yielding a measure of the amplitude of the tethered motion. This ``variance-filtering'' method has the advantage of removing instrumental
drift on time scales slower than the window width, but also degrades the ability to see true dynamics in the bead on time scales faster than the window. 

A better approach might be to include fixed (or ``stuck'') beads in the sample, track them, and subtract their position traces from those of the mobile beads; in practice, however, it is generally not convenient to
prepare such samples. An intermediate method is to track several beads simultaneously, and to use their common mean position as a proxy for the absent fiducial fixed bead position. This ``mean-bead'' 
subtraction method can remove most drift motion, while preserving  information about truly random Brownian motion in the bead traces [\citen{nels06a,beau07a}\cocisstar]. (The residual, unremoved instrumental motion can readily be 
characterized by preparing samples consisting entirely of stuck beads and applying the same analysis to them.)

Unremoved drift can make the distribution of bead positions appear stretched in the direction of drift. Therefore, a final step of bead selection involves assessing the circular symmetry of the 
observed bead positions \textit{after} drift subtraction. Beads whose distributions have principal axes differing by more than 5\% may be attached to the surface by multiple tethers, or be otherwise 
faulty; such traces should be rejected. The surviving beads have equilibrium position distributions that depend only on distance from the anchoring point; \fref{distrs} shows some examples. 

\ifigab{distrs}{Probability distribution functions for equilibrium bead fluctuations.
\textit{Left:}
A typical observed probability distribution function $P(\rho)$ for a 1096 bp-long tether (dots) is distinct from that of a two-dimensional Gaussian with the same mean-square deviation (curve). That is,
the solid curve is the function $(2\rho /\sigma^2) \exp(-(\rho/\sigma)^2)$, where $\sigma = 211\,\nmunit$.
\textit{Right:}
Probability distributions (histograms) $P(\rho)$ for tethers of length $L =$~957, 2211, and 3477 bp (left to right). Dots: For each length, distributions from several tethers are represented in different
colors. Curves: Predictions from theoretical model described in \sref{ef}, with  $\xi = 43\, \nmunit$. 
The bead radius $R_{\rm bead}$ was known to be $240\, \nmunit$ from the manufacturer's specifications.
(Data and theory redrawn after \cite[Figs.~8--9]{nels06a}.)}{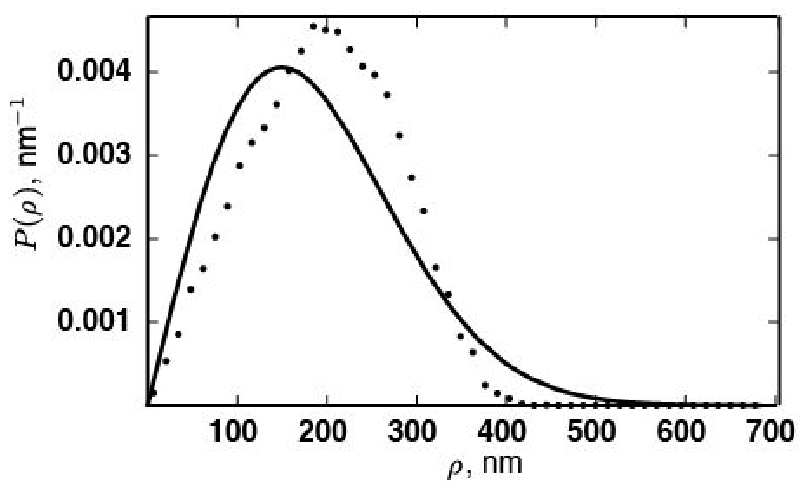}{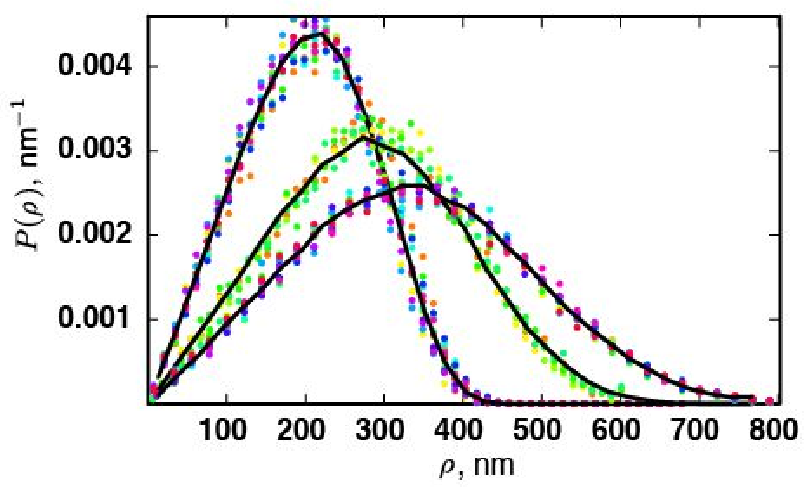}

\section{Theory of tethered Brownian motion}
\subsection{Equilibrium fluctuations\label{s:ef}}
The shapes of the distributions in \fref{distrs} are not Gaussian, and the question arises whether we can predict them from first principles. Beyond such basic polymer science questions, an {\it a priori}
knowledge of, say, the mean-square bead excursion for simple tethers is useful for interpreting experiments where the tether conformation is not known in advance. For example, tether length may be 
changing in time, in a way we would like to measure, as a processive enzyme walks along the DNA. Or, effective tether length may change stochastically as proteins bind to the DNA, kinking or looping 
it. A third motivation for such a theoretical study is that by comparing theory to experiment, we gain confidence both that the experiment is working as desired and that the underlying polymer 
theory, which we may wish to apply to other problems, is adequate.

Although the end-end distribution of a semiflexible polymer such as DNA is a classical problem in polymer physics, the present problem differs from that one in several respects. For example, the
DNA is not isolated, but instead is confined between two surfaces. DNA attached to a single planar surface experiences an effective entropic stretching force due to the steric exclusion from half of
space; a similar effective repulsion exists between the DNA and the large bead. Far more important than these effects, however, is the steric exclusion of the bead from the wall. 
\rref{sega06a}\cocisstar{} argued
that the effect of this exclusion would be to create an entropic stretching force on the DNA of the general form $F_{\rm eff}=\kbt/z$ for $z<2R_{\rm bead}$, where $z$ is the height of the DNA 
endpoint from the wall, $R_{\rm bead}$ is the radius of the bead, and $\kbt$ is the thermal energy at temperature $T$. (For $z>2R_{\rm bead}$ the entropic force is zero.) Even if we do not measure 
$z$ directly, this stretching force affects the distributions of projected distance, reducing its spread. 

Additional subtleties of the problem include the fact that the polymer itself has two additional length scales in addition to the bead radius, namely its persistence length $\xi$ and finite total length $L$, 
and the fact that we do not observe the polymer endpoint, but rather the center of the attached bead. Some of these effects have been studied in an analytical formalism for the low-force case \cite{seol?a}, but 
for \textit{zero} applied stretching force the steric constraints, not fully treatable in that formalism, become important. For this reason \rrefs{sega06a,nels06a} developed a Monte Carlo calculation 
method (a similar method was independently used for a study of DNA cyclization by Czapla et al.~[\citen{czap06a}\cocisstar], and related calculations also appear in \cite{beck05b}).

The Monte Carlo code generated chains realized as strings of rotation matrices, all close to the identity matrix. Each such matrix represents the orientation of one polymer chain segment relative to
its predecessor and was drawn from a Gaussian distribution, as expected for bend angles subject to a harmonic restoring torque \cite{volobook,PhilBook}. The width of the Gaussian bend distribution was
chosen to reproduce the persistence length of DNA, which was either supposed to be known from other experiments in similar solvent conditions, or taken as a fitting parameter. The biotin and
digoxigenin linkages attaching the DNA to bead and wall were treated as freely flexible pivots, and so the orientation of the first chain segment, and that of the bead relative to the last segment,
were taken uniformly distributed in the half-spaces allowed by the respective surfaces. Once a set of relative orientations was selected, the corresponding absolute orientations were found by
successive matrix multiplication. The spatial location of each chain segment was then found by following the 3-axis of each orientation triad, and the chain and bead were then checked for steric
clashes. Chain/bead configurations surviving this check were then entered into histograms of projected 2D distance between bead center and the anchoring point. The corresponding distributions for bead
position are seen in \fref{distrs} to fit the experimental observations quite well after adjusting a single parameter, the chain persistence length $\xi$; the fit value accords with published values
obtained by force--extension measurement.

We can see the trends in the data more clearly if we reduce the distributions to their root-mean-square excursion, a quantity often used in experiments to characterize tethered particle motion.
To facilitate fitting the theory to the data,  the RMS excursion radius was computed for one fixed value of $\xi$ and a
grid of different values of $R_{\rm bead}$ and $L$ in the range of interest, then summarized by an interpolating function. Dimensional analysis shows that $\sqrt{\exv{\rho^2}}/\xi$ can be
written as a dimensionless function of $L/\xi$ and $R_{\rm bead}/\xi$, so this calculation suffices to find the RMS excursion at any $\xi,\  L,$ and $R_{\rm bead}$ \cite{nels06a}.  The resulting 
interpolating function can then be readily fit to experimental data. 

\ifig{calibrplot}{First-principles prediction of equilibrium bead excursion.
\textit{Dots:} Experimental values for RMS motion of bead center for the three different bead sizes: Top to bottom, $R_{\rm bead}=485,\ 245,$ and $100\,\nmunit$. Each dot represents the 
average of approximately 20--200 different observed beads with the given tether length. \cut{Error bars represent DESCRIBE ERROR BARS HERE.}
\textit{Curves:} Theoretically predicted RMS motion, corrected for the blurring effect of our long shutter time. The solid curves assume persistence length $\xi=38\,\nmunit$; the dashed curves assume 
$\xi=45\,\nmunit$. 
There are \textit{no other fit parameters;} the theoretical model uses values for bead diameter given by the manufacturer's specification. (Data and theory from Lin Han et al., unpublished data
.)
}{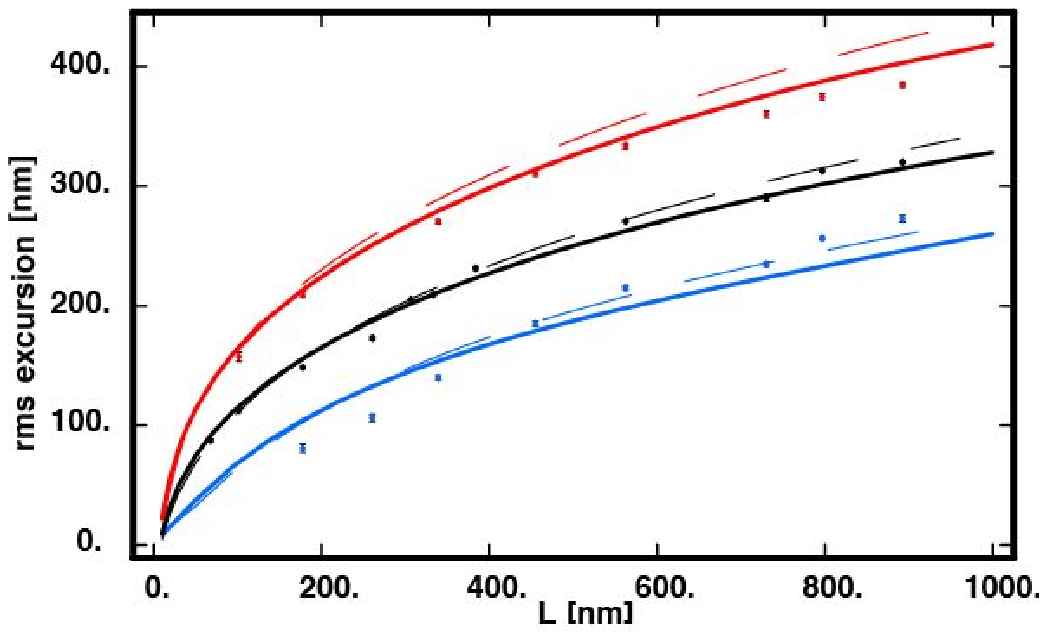}
The above procedure works well for TPM data taken with a fast shutter speed \cite{nels06a}. For longer exposure times, however, the theory must be corrected to account for the blurring of the image due to 
Brownian motion during each exposure (Lin Han et al., unpublished data)
. \fref{calibrplot} shows a global fit of the corrected theory to excursion data for three different bead radii. The discrepancy between theory and 
experiment may reflect unremoved instrumental drift, for example high-frequency motion that the filtering algorithm could not distinguish 
from true Brownian motion. Alternatively, the effective bead size may be slightly different from the nominal value, or effectively different due to surface irregularity. Finally, hydrodynamic effects slowing
 the bead's motion when close to the surface may set an equilibration time scale longer than the observation time.

\subsection{Dynamics of tethered particle motion}
The previous subsection discussed {equilibrium} bead fluctuations, but for the study of dynamic phenomena we also wish to understand the \textit{dynamics} of tethered motion. A simple model captures 
qualitatively the expected character of the motion \cite{beau07b}. We imagine dividing time into very small intervals. At each time step, the change in bead position relative to the previous step is 
a combination of a random, diffusive step (Brownian motion), plus a deterministic drift step determined by the particle's current position in an effective potential well. The potential well comes 
from the entropic elasticity of the tether. The friction constant for the deterministic component of the motion is related to the diffusion constant for the Brownian component by the Einstein relation.
If we roughly model the potential as that of a Gaussian chain, then the problem becomes the fluctuations of an overdamped harmonic oscillator, which can be 
solved analytically for the RMS excursion and the temporal autocorrelation function \cite{Breif65a}. 

Indeed, experimental data for the equilibrium excursion distribution and the 2D autocorrelation function conform to the qualitative forms expected from the above description (namely the Gaussian
distribution and exponential falloff in autocorrelation respectively) \cite{beau07b}. However, for more detailed comparison of theory to experiment we would need to account for many modifications to
the simple picture. For example, the friction and diffusion constants are much bigger than the values from Stokes's law, due to wall-proximity effects. Nor is the motion in the third (height) 
direction decoupled from projected-plane motion, as assumed in the simple model above. Nor is a DNA tether's entropic elasticity well represented by the
Gaussian-chain formula. For all these reasons, prediction of TPM dynamics from first principles is a daunting task. But we can at least be fairly confident that whatever this motion, on video time 
scales it will have a Markov property: Each step will be drawn from a distribution depending only on the position at the start of that step (not on previous steps). \rref{beau07a
}\cocisstar{} and (J. F. Beausang and P. C. Nelson, unpublished data) extracted 
the single-step probability distributions from pairs of adjacent video frames, found a convenient parametric representation for these empirical functions, and then showed that they could be used to 
generate simulated TPM data with the same equilibrium and dynamics over long times as the real data. Knowing the dynamics of tethers that are not undergoing discrete conformational changes sets the 
stage for investigating tethers that \textit{are} changing, for example by loop formation.

\section{DNA looping}
\subsection{Background}
All cells  use genetic regulatory mechanisms that involve, among other things, the formation of loops in DNA similar to the one shown in \fref{cartoon}. Particularly well studied cases 
involve bacteria, for example the lactose metabolism and lambda phage genetic switches
\cite{ripp95a,matt98a,reve99a,saiz06b}. Reconstituting DNA looping behavior \textit{in vitro}, and studying its dependence on physical parameters of the system, is thus an important step in clarifying the mechanism 
of gene regulation. A particular puzzling aspect of regulation by looping is that even very short loops, corresponding to tightly bent DNA, seem to form readily. In some cases, DNA-bending proteins 
appear to assist in loop formation; in others, specially flexible or pre-bent DNA sequence may help too (e.g. see \cite{beck05a}). But a number of experiments appear to point to a surprising rate of loop formation even in the 
absence of such assistance. 

\subsection{TPM measurements of looping}
TPM experiments are beginning to shed new light on looping [\citen{finz95a}\cocisstarstar,\citen{zurl06a},\citen{broe06a}\cocisstarstar,\citen{norm07a},\citen{vanz06a},\citen{beau07a
}\cocisstar] (O. K. Wong, M. Guthold, D. A. Erie, J. Gelles, personal communication; J. F. Beausang and P. C. Nelson, unpublished data). These experiments appear to be exempt from some of the 
complexities in corresponding DNA cyclization experiments. Moreover, by watching repeated loop formation and breakdown in real time, they can yield not only equilibrium constants but also 
corresponding rates, allowing us to quantify separately the kinetic barriers to loop formation (such as DNA bending energy) and breakdown (controlled by the unbinding kinetics of regulatory 
proteins and their operator sequences in the DNA) \cite{vanz06a}. Finally, the TPM technique allows us to monitor a single molecule continuously as we change its surroundings, for example before and 
after the addition of the regulatory proteins [\citen{beau07a}\cocisstar]. 

Looping events are sometimes clearly discernible by eye in the microscopy, and in any case can be seen as sharp transitions in the time series after drift subtraction (\fref{FIGTIMERHO}).  But a 
glance at the figure makes it clear that the interesting looping transitions are partially obscured by the bead's Brownian motion. \sref{dk} below will outline some new statistical methods to draw 
conclusions about looping from such data.

\section{Theoretical models of DNA looping}
\subsection{Equilibrium calculation of looping $J$ factor}
Before discussing looping kinetics, however, we again begin with equilibrium. Classical works pointed out that the equilibrium constant for looping, or for cyclization (ring formation),
can usefully be decomposed into \textit{(i)}~a binding constant for free operator DNA fragments to bind to the repressor protein, and \textit{(ii)}~a ``$J$ factor'' contribution, 
describing the effect of the DNA chain joining the two operators \cite{jaco50a,yoon74a,sute76a,mark82a}. The $J$ factor can roughly be regarded as the concentration of one operator in the vicinity of the 
other. Its overall dependence on the length  of the intervening DNA between the operators can be qualitatively understood as reflecting  two competing phenomena. First, a shorter tether confines the 
second operator into a smaller region about the first one, increasing the effective concentration. But if the required loop is too short, then forming it will entail a large bending elastic energy 
cost, depressing the probability by a Boltzmann factor. For these reasons, the cyclization $J$ factor exhibits a peak at DNA length about 300 basepairs \cite{shim84a}. Later work extended Shimada and 
Yamakawa's calculation in many ways, using a variety of mathematical techniques
[\citen{shim84a,hage85a,leve86a,podt00a,zhan03a,spak04a,spak05a,yan05a,doua05a,zhan06a,wigg06a,spak06a},\citen{czap06a}\cocisstar]. Recent work has attempted to extract the looping $J$ factor from \textit{in vivo} 
experimental data \cite{bint05b,saiz05a}, but any such determination contains uncertainties due to the complex world inside a living cell. 

To clarify the situation, other experiments cited earlier 
have sought to quantify looping \textit{in vitro}. Confronting these experiments with theory has required new calculations to incorporate the specific features of TPM.
%
%
For example, shortening the DNA construct increases the 
entropic force exerted by bead--wall avoidance, discouraging looping \cite{sega06a}.
To see what a measurement of TPM looping equilibrium tells us, we can calculate
the expected local concentration of one operator near the other, based on a particular mathematical model of DNA elasticity (Lin Han et al., unpublished data). Those authors chose a harmonic-elasticity model (a generalization of the 
traditional wormlike chain model), to see if it could adequately explain their experimental results, or if, on the contrary, some non-harmonic model (for example the one proposed in 
[\citen{yan04a}\cocisstar,\citen{wigg05a}\cocisstar]) might be indicated.

As in \sref{ef} above, the nonlocal steric constraints dictated a Monte Carlo approach. L.~Han et al.
modified the Gaussian sampling method outlined in \sref{ef}
to generate many simulated DNA chains,  apply
steric constraints \cite{sega06a}, and report what fraction of accepted chain/bead configurations had the two operator sites separated by 7$\,\nmunit$ (the distance 
between operator centers in the LacR--DNA complex, as seen in Protein Data Bank structure 1LBG). The standard elastic model as an isotropic rod is inadequate for the description of DNA loops only a few helical repeats in length 
(see for instance [\citen{czap06a}\cocisstar]), so the authors modified the elasticity model to account for bend anisotropy and bend--roll coupling. 
The chain generation accounted for bead--wall, bead--chain, and wall--chain avoidance, but not chain--chain, which is much less important for the short tethers used in TPM experiments.

The result of the simulation was that the looping $J$ factor for short loops was about 0.01 times as great for the constructs with interoperator spacing around 100$\,$bp as for those with 
spacing around 300$\,$bp; this ratio was more than a hundred times smaller than the experimentally determined ratio of 1.7 at optimum helical phasing (Lin Han et al., unpublished data)
. The authors concluded that the hypotheses of
harmonic elasticity, plus a rigid V-shaped protein coupler, could not explain their results. One possible explanation, 
for which other support has been growing, is the alternate hypothesis of DNA elastic breakdown at high curvature [\citen{yan04a}\cocisstar,\citen{wigg05a}\cocisstar].

\section{Determination of kinetics and event detection by hidden Markov modeling\label{s:dk}}
The experimental determination of the fraction of time spent in the looped state seems to require that we know the looping state of 
the DNA tether at every moment of time. Indeed if that information were known, then we could also find  the kinetic rate constants mentioned earlier, by tabulating the dwell times for loops to form 
and break down and then fitting them to exponential (or multiexponential) distributions. Unfortunately, data such as those in \fref{FIGTIMERHO} do not admit an obvious point-by-point determination of 
looping state, in part because the probability distributions of bead positions in the different looped states have large overlap (\fref{distrs}).

One popular approach to this problem has been to run a sliding window across the data, typically of 4$\,\sunit$ width, and within the window to apply the variance filter  mentioned in \sref{tt}. The 
filtered signal is  local measure of the amplitude of the tethered Brownian motion; loop formation and breakdown events are defined as time points where it crosses a predetermined threshold 
\cite{vanz06a}, and then dwell time histograms are generated. There are several difficulties with this window/threshold approach, however [\citen{beau07a
}\cocisstar], including a strong dependence of the 
answers on the choice of window size. One approach to (one part of) this problem is to correct for missed transitions \cite{vanz06a}.
An alternative method, generally useful for obtaining kinetic and equilibrium constants from noisy data, is hidden Markov modeling (HMM) [\citen{rabi89a}\cocisstarstar], which uses no filter at all. 

The key point in HMM is the realization that it is not necessary to identify unambiguously the states at each time point. Instead, a kinetic model is assumed for the underlying ``hidden'' Markov process of 
interest (here the loop/unloop transitions). The model contains a small number of unknown parameters (here, the rate constants $\vec\alpha=(\tau_{\rm LF},\tau_{\rm LB})$ 
for loop formation and breakdown). The model also assumes that given the 
underlying hidden state $q$, the probability distribution of the observed quantity (here the bead position $\vr$) is known. That is, we choose a transition matrix $T(q_{n+1}|q_{n}; \vec\alpha)$ and 
measure the conditional probabilities $P(\vr|q)$, for example by studying DNA tethers that are forced to be permanently looped or unlooped.

If we knew the exact states of the hidden variable at all times, $\{q_i,i=1,\ldots,N\}$, then we could find the likelihood, or the probability that a particular time series $\cal O$ of the observed quantity,
${\cal O}=\{\vr_i,i=1,\ldots,N\}$, would arise. This likelihood contains contributions from both the transition matrix and the conditional probabilities. Since we \textit{don't} know the underlying hidden 
variable, we marginalize it by summing the likelihood over \textit{all} possible time series $\{q_i\}$. Substituting an actual experimentally observed time series $\{\vr_i\}$ gives its likelihood to 
have been generated from the underlying model, as a function of the model's parameters, the unknown rates $\vec\alpha$. If the available time series is very long, we expect that this likelihood 
$P^*({\cal O};\vec\alpha)$ will be very sharply peaked about the true values of the rate constants $\vec\alpha$. That is, our best guess for the system's rate constants is the set of ``maximum likelihood'' 
values.

\ifig{FIGLOGLIK}{Log-likelihood surface for loop formation and breakdown times.
Evaluation of $\log[P^*({\cal O};\vec\alpha)]$ on a logarithmically-spaced grid of $\tlf, \tlb$ lifetimes corresponding to data from \fref{FIGTIMERHO}. (Data and theory from 
J. F. Beausang and P. C. Nelson, unpublished data.)}{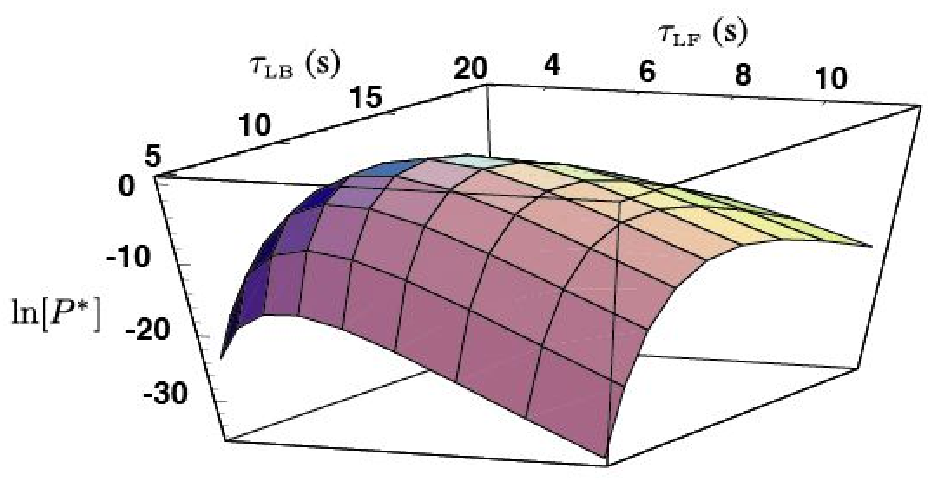}
Hidden Markov analysis has recently been brought to bear on photon-counting single molecule techniques such as FRET
[\citen{andr03a}\cocisstar,\citen{mcki06a}\cocisstar]. To adapt it to TPM experiments requires a slight generalization, however [\citen{beau07a
}\cocisstar], because the observed bead position, being a form of Brownian motion, has some 
memory. Indeed, both the observed motion and also the underlying hidden state have transition matrices describing the Markov character of their time evolution. Nevertheless, we can measure 
experimentally the transition matrix for the observed variable $\vr$, using control experiments, and so reduce our problem to maximizing a likelihood that depends only on two unknown rate constants 
for the hidden state transitions. The resulting function generally displays a single smooth peak (\fref{FIGLOGLIK}), and so can be maximized numerically by standard methods.

\section{Outlook}
Tethered particle motion experiments are both reasonably simple to implement, and reasonably simple to interpret. This review has focused on illustrative results showing that straightforward,
computationally inexpensive Monte Carlo simulation gives a good account of both equilibrium and dynamic properties of simple tethered motion. Such agreement lets us use TPM as a test bench for
studying more biophysically interesting processes, such as DNA looping and processive enzyme dynamics.

\section*{Acknowledgements} 
I thank Nily Dan and Yale Goldman for many technical discussions related to the work described here, as well as my coauthors
John Beausang,
David Dunlap, 
Laura Finzi,
Lin Han,
Hernan Garcia,
Rob Phillips,
Kevin Towles,
and Chiara Zurla.
This work was partially supported by
NSF Grants DMR-0404674 and DGE-0221664, and the Penn Nano/Bio Interface
Center (NSF DMR04-25780).


\bibliographystyle{unsrt}
\bibliography{BIBboulder}

\end{document}